\documentclass[usenatbib, longauth,letter]{aa} 

\usepackage{graphicx}
\usepackage{graphics}
\usepackage{txfonts}
\usepackage{epsfig}
\usepackage{natbib}
\usepackage[]{color}

\begin{document}

\title{Detection of the gravitational redshift in the orbit of the star S2
       near the Galactic centre massive black hole}

\titlerunning{Detection of gravitational redshift}
\subtitle{}

\author{GRAVITY Collaboration\thanks{GRAVITY is developed
    in a collaboration by the Max Planck Institute for
    extraterrestrial Physics, LESIA of Paris Observatory / CNRS / Sorbonne Universit\'e
    / Univ. Paris Diderot and IPAG of Universit\'e Grenoble Alpes /
    CNRS, the Max Planck Institute for Astronomy, the University of
    Cologne, the CENTRA - Centro de Astrofisica e Gravita\c c\~ao, and
    the European Southern Observatory. Corresponding author:
    F.~Eisenhauer (eisenhau@mpe.mpg.de)}:
R.~Abuter\inst{8}
\and A.~Amorim\inst{6,14}
\and N.~Anugu\inst{7}
\and M.~Baub\"ock\inst{1}
\and M.~Benisty\inst{5}
\and J.P.~Berger\inst{5,8}
\and N.~Blind\inst{10}
\and H.~Bonnet\inst{8}
\and W.~Brandner\inst{3}
\and A.~Buron\inst{1} 
\and C.~Collin\inst{2}
\and F.~Chapron\inst{2}
\and Y.~Cl\'{e}net\inst{2}
\and V.~Coud\'e~du~Foresto\inst{2}
\and P.T.~de~Zeeuw\inst{12,1}
\and C.~Deen\inst{1}
\and F.~Delplancke-Str\"obele\inst{8}
\and R.~Dembet\inst{8,2}
\and J.~Dexter\inst{1}
\and G.~Duvert\inst{5}
\and A.~Eckart\inst{4,11}
\and F.~Eisenhauer\inst{1}
\and G.~Finger\inst{8}
\and N.M.~Förster~Schreiber\inst{1} 
\and P.~Fédou\inst{2}
\and P.~Garcia\inst{7,14}
\and R.~Garcia~Lopez\inst{16,3}
\and F.~Gao\inst{1}
\and E.~Gendron\inst{2}
\and R.~Genzel\inst{1,13}
\and S.~Gillessen\inst{1}
\and P.~Gordo\inst{6,14} 
\and M.~Habibi\inst{1}
\and X.~Haubois\inst{9}
\and M.~Haug\inst{8} 
\and F.~Haußmann\inst{1}
\and Th.~Henning\inst{3}
\and S.~Hippler\inst{3}
\and M.~Horrobin\inst{4}
\and Z.~Hubert\inst{2,3} 
\and N.~Hubin\inst{8}
\and A.~Jimenez~Rosales\inst{1}
\and L.~Jochum\inst{8}
\and L.~Jocou\inst{5}
\and A.~Kaufer\inst{9}
\and S.~Kellner\inst{11} 
\and S.~Kendrew\inst{15,3}
\and P.~Kervella\inst{2}
\and Y.~Kok\inst{1}
\and M.~Kulas\inst{3} 
\and S.~Lacour\inst{2}
\and V.~Lapeyr\`ere\inst{2}
\and B.~Lazareff\inst{5}
\and J.-B.~Le~Bouquin\inst{5}
\and P.~L\'ena\inst{2}
\and M.~Lippa\inst{1}
\and R.~Lenzen\inst{3} 
\and A.~M\'erand\inst{8}
\and E.~Müller\inst{8,3} 
\and U.~Neumann\inst{3} 
\and T.~Ott\inst{1}
\and L.~Palanca\inst{9} 
\and T.~Paumard\inst{2}
\and L.~Pasquini\inst{8}
\and K.~Perraut\inst{5}
\and G.~Perrin\inst{2}
\and O.~Pfuhl\inst{1}
\and P.M.~Plewa\inst{1}
\and S.~Rabien\inst{1}
\and A.~Ram\'irez\inst{9} 
\and J.~Ramos\inst{3} 
\and C.~Rau\inst{1} 
\and G.~Rodr\'iguez-Coira\inst{2}
\and R.-R.~Rohloff\inst{3} 
\and G.~Rousset\inst{2}
\and J.~Sanchez-Bermudez\inst{9,3}
\and S.~Scheithauer\inst{3}
\and M.~Sch\"oller\inst{8}
\and N.~Schuler\inst{9} 
\and J.~Spyromilio\inst{8}
\and O.~Straub\inst{2}
\and C.~Straubmeier\inst{4}
\and E.~Sturm\inst{1}
\and L.J.~Tacconi\inst{1}
\and K.R.W.~Tristram\inst{9}
\and F.~Vincent\inst{2}
\and S.~von~Fellenberg\inst{1}
\and I.~Wank\inst{4} 
\and I.~Waisberg\inst{1}
\and F.~Widmann\inst{1}
\and E.~Wieprecht\inst{1}
\and M.~Wiest\inst{4}
\and E.~Wiezorrek\inst{1} 
\and J.~Woillez\inst{8}
\and S.~Yazici\inst{1,4}
\and D.~Ziegler\inst{2}
\and G.~Zins\inst{9}
}

\institute{
Max Planck Institute for extraterrestrial Physics,
Giessenbachstraße~1, 85748 Garching, Germany
\and LESIA, Observatoire de Paris, Universit\'e PSL, 
CNRS, Sorbonne Universit\'e, Univ. Paris Diderot, 
Sorbonne Paris Cit\'e, 5 place Jules Janssen, 92195 Meudon, France
\and Max Planck Institute for Astronomy, K\"onigstuhl 17, 
69117 Heidelberg, Germany
\and $1^{\rm st}$ Institute of Physics, University of Cologne,
Z\"ulpicher Straße 77, 50937 Cologne, Germany
\and Univ. Grenoble Alpes, CNRS, IPAG, 38000 Grenoble, France
\and Universidade de Lisboa - Faculdade de Ci\^encias, Campo Grande,
1749-016 Lisboa, Portugal 
\and Faculdade de Engenharia, Universidade do Porto, rua Dr. Roberto
Frias, 4200-465 Porto, Portugal 
\and European Southern Observatory, Karl-Schwarzschild-Straße 2, 85748
Garching, Germany
\and European Southern Observatory, Casilla 19001, Santiago 19, Chile
\and Observatoire de Gen\`eve, Universit\'e de Genève, 51 Ch. des
Maillettes, 1290 Versoix, Switzerland
\and Max Planck Institute for Radio Astronomy, Auf dem H\"ugel 69, 53121
Bonn, Germany
\and Sterrewacht Leiden, Leiden University, Postbus 9513, 2300 RA
Leiden, The Netherlands
\and Departments of Physics and Astronomy, Le Conte Hall, University
of California, Berkeley, CA 94720, USA
\and CENTRA - Centro de Astrof\'{\i}sica e
Gravita\c c\~ao, IST, Universidade de Lisboa, 1049-001 Lisboa,
Portugal
\and European Space Agency, Space Telescope Science Institute, 
\mbox{3700 San Martin Drive}, Baltimore MD 21218, USA
\and Dublin Institute for Advanced Studies, 
31 Fitzwilliam Place, \mbox{Dublin 2}, Ireland
}

\date{{\bf This paper is dedicated to Tal Alexander, who passed away about a week before the pericentre approach of S2 } 
\newline \newline 
Accepted for publication in A\&A Letters, 29 June 2018}

\abstract{The highly elliptical, 16-year-period orbit of the star S2 around the massive black hole candidate Sgr\,A* is a sensitive probe of the gravitational field in the Galactic centre. Near pericentre at $120\,\mathrm{AU}\,{\approx}\,1400$ Schwarzschild radii, the star has an orbital speed of ${\approx}\,7650\,\mathrm{km/s}$, such that the first-order effects of Special and General Relativity have now become detectable with current capabilities. Over the past 26 years, we have monitored the radial velocity and motion on the sky of S2, mainly with the SINFONI and NACO adaptive optics instruments on the ESO Very Large Telescope, and since 2016 and leading up to the pericentre approach in May 2018, with the four-telescope interferometric beam-combiner instrument GRAVITY. From data up to and including pericentre, we robustly detect the combined gravitational redshift and relativistic transverse Doppler effect for S2 of $z\!=\!\Delta\lambda / \lambda\,{\approx}\,200\,\mathrm{km/s}/c$ with different statistical analysis methods. When parameterising the post-Newtonian contribution from these effects by a factor $f$, with $f\!=\!0$ and $f\!=\!1$ corresponding to the Newtonian and general relativistic limits, respectively, we find from posterior fitting with different weighting schemes $f\!=\!0.90\!\pm\!0.09|_\mathrm{stat}\!\pm\!0.15|_\mathrm{sys}$. The S2 data are inconsistent with pure Newtonian dynamics.
}


\maketitle

\section{Introduction}
\label{sec:intro}

General Relativity (GR) so far has passed all experimental tests with flying colours  (\citealt{1916AnP...354..769E, 2014LRR....17....4W}). The most stringent are tests that employ solar mass pulsars in binary systems \citep{2006Sci...314...97K}, and gravitational waves from $10\!-\!30\,M_\odot$ black hole in-spiral events \citep{2016ApJ...818L..22A,2016PhRvL.116f1102A, 2016PhRvL.116x1103A}. These tests cover a wide range of field strengths and include the strong curvature limit (Fig.~\ref{fig:a2}). At much lower field strength, Earth laboratories probe planetary masses that are about a factor $10^6$ lower than the stellar mass scale. For massive black hole (MBH) candidates with masses of $10^{6\!-\!10}\,M_\odot$ , only indirect evidence for GR effects has been reported, such as relativistically broadened, redshifted iron K$\alpha$ line emission in nearby active galaxies \citep{1995Natur.375..659T, 2000PASP..112.1145F}. The closest MBH is at the centre of the Milky Way ($R_0\,{\approx}\,8\,$kpc, $M_\bullet\,{\approx}\,4\!\times\!10^6\,M_\odot$), and its Schwarzschild radius subtends the largest angle on the sky of all known MBHs ($R_S\,{\approx}\,10\,\mu{\mathrm as}\,{\approx}\,0.08\,\mathrm{AU}$). It is coincident with a very compact, variable X-ray, infrared, and radio source, Sgr\,A*, which in turn is surrounded by a very dense cluster of orbiting young and old stars. Radio and infrared observations have provided detailed information on the distribution, kinematics, and physical properties of this nuclear star cluster and of the hot, warm, and cold interstellar gas interspersed in it (cf. \citealt{2010RvMP...82.3121G, 2012RAA....12..995M, 2013CQGra..30x4003F}). High-resolution near-infrared (NIR) speckle and adaptive optics (AO) assisted imaging and spectroscopy of the nuclear star cluster over the past 26 years, mainly by two groups in Europe (the Max Planck Institute for Extraterrestrial Physics, MPE, and the University of Cologne at the ESO New Technology Telecsope, NTT, and the Very Large Telescope, VLT) and one group in the USA (the University of California at Los Angeles, UCLA, at the Keck telescopes) have delivered more than $10^4$ stellar motions and orbit determinations for ${\approx}\,45$ individual stars \citep{2002Natur.419..694S, 2003ApJ...586L.127G, 2008ApJ...689.1044G, 2005ApJ...628..246E, 2009ApJ...692.1075G, 2017ApJ...837...30G, 2009A&A...502...91S, 2012Sci...338...84M, 2016ApJ...830...17B, 2016ApJ...821...44F}. These orbits, in particular, the highly eccentric orbit of the main-sequence B-star S2 (or S02 in the UCLA nomenclature), have demonstrated that the gravitational potential is dominated by a compact object of ${\approx}\,4\!\times\!10^6\,M_\odot$ that is concentrated within a pericentre distance from S2 of 17 light hours ${\approx}\,14\,\mathrm{mas}$ or $120\,\mathrm{AU}$ from Sgr\,A*. S2 appears to be a single star \citep{2008ApJ...672L.119M, 2017ApJ...847..120H, 2017A&A...602A..94G, 2018ApJ...854...12C}, making it an ideal probe for testing GR by diffraction-limited imaging and spectroscopy \citep{2005PhR...419...65A, 2006ApJ...639L..21Z, 2017ApJ...845...22P}, and interferometry \citep{2017A&A...608A..60G} through the deviation of its apparent motion from a Keplerian orbit. 

The radio source Sgr\,A* is coincident with the mass centroid to ${<}\,1\,\mathrm{mas}$ \citep{2015MNRAS.453.3234P}, and is itself very compact, with $R\,(1.3\,\mathrm{mm})\,{<}\,18\,\mu{\mathrm as}\,{\approx}\,1.8\,R_S$, based on millimetre very long baseline interferometry \citep{2000ApJ...528L..13F, 2008Natur.455...78D, 2017ApJ...850..172J}. In addition, Sgr\,A* shows no detectable intrinsic motion, which supports the interpretation that the compact radio source is coincident with the mass \citep{2004ApJ...616..872R, 2009IJMPD..18..889R}. The most conservative explanation for Sgr\,A* is that it is an MBH, assuming that GR is applicable \citep{2010RvMP...82.3121G, 2013CQGra..30x4003F,2016CQGra..33j5015V}. So far, Newtonian orbits in a single central force potential can describe the motions of all stars. Any extended mass within the S2 orbit is lower than about $1\,$\% of the central mass \citep{2017PhRvL.118u1101H, 2017ApJ...837...30G}.

\section{Observations}
\label{sec:observations}

\begin{figure}[t!]
\vspace{-4.3cm}
\hspace{-0.8cm}
\includegraphics[width=11.8cm]{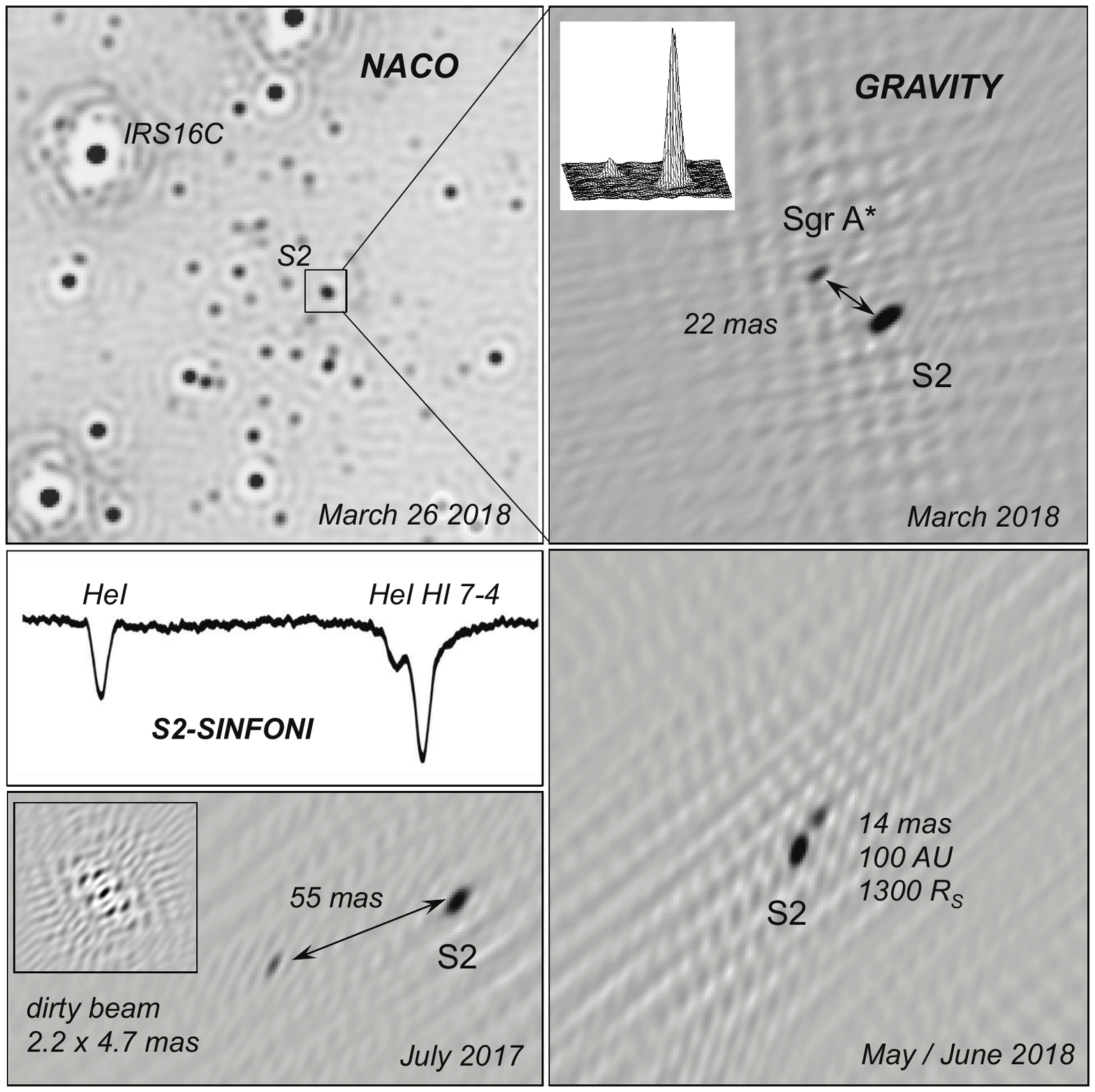}
\vspace{-3.8cm}
\caption{Monitoring the S2 orbit around Sgr\,A* with the three VLT(I) instruments NACO (AO-assisted, single UT imaging), GRAVITY (interferometric astrometry-imaging with all four UTs of the VLT) and SINFONI (AO-assisted integral field spectroscopy). Upper left: Deconvolved NACO K-band image of the Galactic centre a few weeks before the 2018 pericentre passage. The source S2 appears slightly elongated because of confusion with Sgr\,A*. Upper right: Nearly simultaneous GRAVITY image of S2 and Sgr\,A*. The image shows the central 150 mas ($0.0059\,\mathrm{pc}\,{\approx}\,1.4 \times 10^4\,R_S$) after self-calibration and CLEANing with AIPS. The image is reconstructed from 34 integrations of 5 minutes each from several nights at the end of March 2018. Sgr\,A* was (on average) $K\!=\!16.6\,$mag and the rms noise level of the background after cleaning is  ${\approx}\,20\,$mag. Both Sgr\,A* and S2 are unresolved. Here and in the other GRAVITY images, the elongation  is due to the shape of the interferometric clean beam. Bottom: S2~-~Sgr\,A* GRAVITY images (co-addition of several days) from July 2017 (bottom left) and May\,/\,June 2018, a few days after pericentre (bottom right). The inset in the bottom left panel shows the instantaneous interferometric beam, which is $2.2\,\mathrm{mas} \times 4.7\,\mathrm{mas}$ without Earth rotation. The inset in the middle left shows a co-added SINFONI K-band spectrum of the star S2, taken from \citet{2017ApJ...847..120H}.}
\label{fig:fig1}
\end{figure}

\begin{figure*}[t!]
\centering
\vspace{-0.7cm}
\includegraphics[width=20cm]{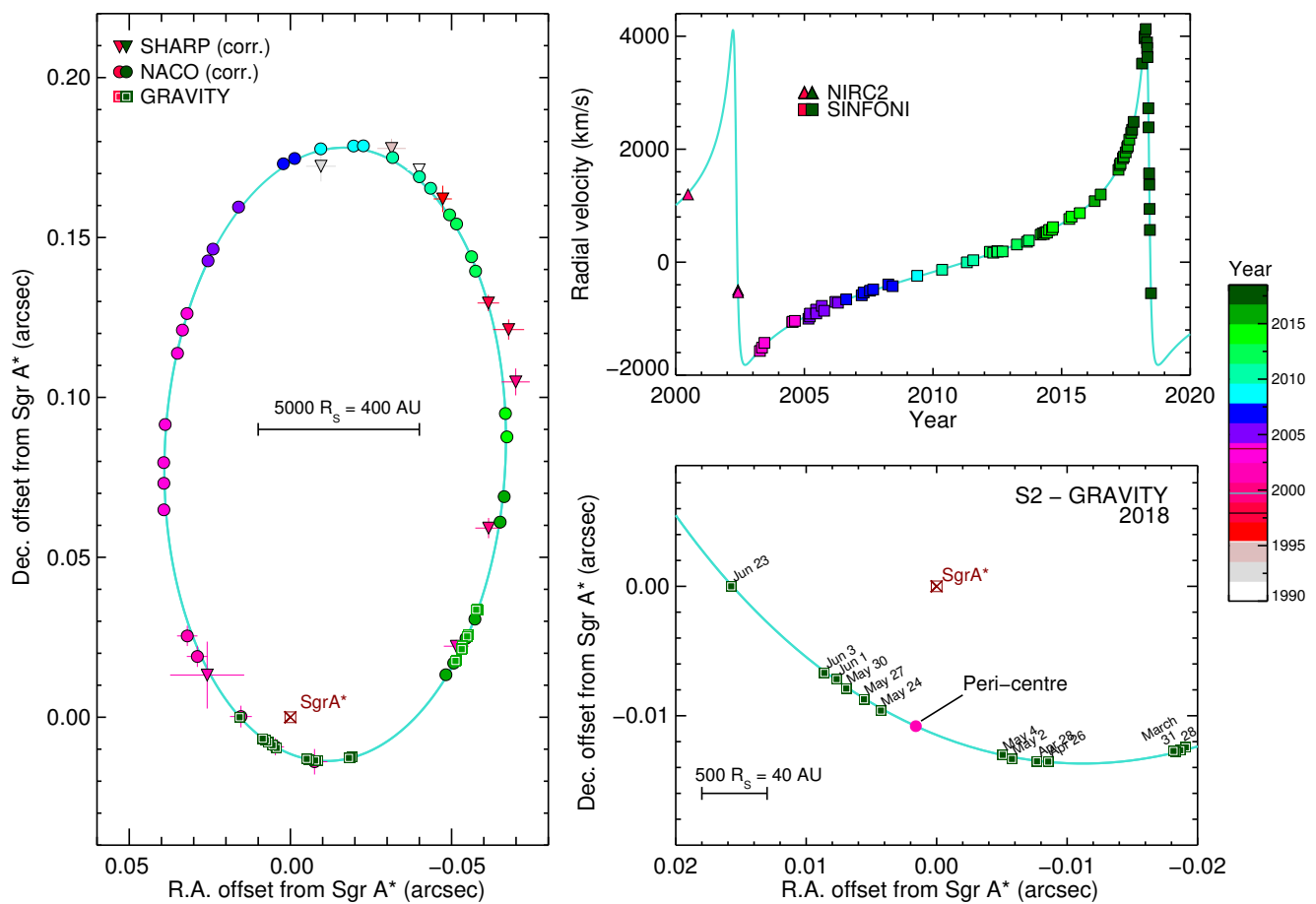}
\vspace{-3cm}
\caption{Summary of the observational results of monitoring the S2~-~Sgr\,A* orbit from 1992 to 2018. Left: Projected orbit of the star S2 on the sky (J2000) relative to the position of the compact radio source Sgr\,A* (brown crossed square at the origin). Triangles and circles (and $1\,\sigma$ uncertainties) denote the position measurements with SHARP at the NTT and NACO at the VLT, colour-coded for time (colour bar on the right side). All data points are corrected for the best-fit zero-point $(x_0, y_0)$ and drifts $(\dot{x}_0,\dot{y}_0)$ of the coordinate system relative to Sgr\,A* (see \citealt{2015MNRAS.453.3234P}). Green squares mark the GRAVITY measurements. The bottom right panel shows a zoom around pericentre in 2018. Top right: Radial velocity of S2 as a function of time (squares: SINFONI/NACO at the VLT; triangles: NIRC2 at Keck). S2 reached pericentre of its orbit at the end of April 2002, and then again on May 19\textsuperscript{th}, 2018 (MJD 58257.67). The data before 2017 are taken from \citet{2008ApJ...689.1044G}, \citet{2016ApJ...830...17B}, \citet{2018ApJ...854...12C}, and \citet{2009ApJ...692.1075G, 2017ApJ...837...30G}. The 2017/2018 NACO/SINFONI and GRAVITY data are presented here for the first time. The cyan curve shows the best-fitting S2 orbit to all these data, including the effects of General and Special Relativity. }
\label{fig:fig2}
\end{figure*}

We present\footnote{Based on observations made with ESO Telescopes at the La Silla Paranal Observatory under programme IDs 075.B-0547, \mbox{076.B-0259}, 077.B-0014, 078.B-0136, 179.B-0261, 183.B-0100, 087.B-0117, 088.B-0308, 288.B-5040, 089.B-0162, 091.B-0081, 091.B-0086, 091.B-0088, 092.B-0238, 092.B-0398, 093.B-0217, 093.B-0218, 594.B-0498, 097.B-0050, 598.B-0043, 299.B-5014, 299.B-5056, 099.B-0162, 0100.B-0731, 0101.B-0195, and 0101.B-0576.} an analysis of the positions and K-band spectra of the star S2 from 1992 to 2018 (Figs.~\ref{fig:fig1} and~\ref{fig:fig2}). 

We obtained sky-projected positions of the star S2 with the speckle camera SHARP at the NTT (1992-2002: \citealt{1993Ap&SS.205....1H}), but most of our imaging comes from the AO-assisted NIR imager NACO at the VLT (2002-2018: \citealt{1998SPIE.3354..606L, 1998SPIE.3353..508R}) and the interferometric astrometry-imager GRAVITY with all four Unit Telescopes (UTs) of the VLT interferometer  \citep{2017A&A...602A..94G}. The SHARP/NACO data deliver relative positions between stars in the nuclear star cluster. These are then registered with $\leq\!1\,\mathrm{mas}$ precision in the radio frame of the Galactic centre \citep{2007ApJ...659..378R} using multi-epoch observations of nine SiO maser stars common between our infrared data and the radio interferometry, after correcting NACO image distortions with observations of a globular cluster calibrated on data from the Hubble Space Telescope \citep{2015MNRAS.453.3234P}. In the GRAVITY interferometric observations, we detected and stabilised the interferometric fringes on the stars IRS16C or IRS16NW located ${\approx}\,1''$ from Sgr\,A*, and observed the ``binary'' S2~-~Sgr\,A* within the second phase-referenced fibre (see \citealt{2017A&A...602A..94G}). S2 and Sgr\,A* are simultaneously detected as two unresolved sources in ${>}\,90\,$\% of our individual integrations (5 minutes each), such that the S2~-~Sgr\,A* vector is directly obtained in each of these measurements.

Our 2003-2018 measurements of the \mbox{Brackett-$\gamma$} line velocity were taken with the AO-assisted integral field spectrometer SINFONI at the VLT \citep{2003SPIE.4841.1548E, 2004Msngr.117...17B}, with five additional 2000-2003 slit-spectra from the AO imagers and spectrometers NIRC2 at Keck (see \citealt{2003ApJ...586L.127G, 2018ApJ...854...12C}) and NACO \citep{2003ApJ...597L.121E}.

The 1992-2016 speckle and AO-imaging and spectroscopic data used below have been presented in \citet{2017ApJ...837...30G}. In 2017 and 2018 we increased the cadence of the observations in preparation for the pericentre approach in May 2018. We added 21 epochs of NACO K- and H-band imaging in the 13~mas/pix scale, and 2 epochs of NACO K-band imaging in the 27~mas/pix scale to measure the SiO maser positions \citep{2007ApJ...659..378R} that define our coordinate system \citep{2015MNRAS.453.3234P}. We obtained 30 data sets of GRAVITY interferometry, and 26 additional spectroscopy epochs with SINFONI using the 25\,mas/pix scale and the combined H+K-band grating with a spectral resolution of $R{\approx}\,1500$. 

For more details on the data analysis of all three instruments, we refer to Appendix A.

\section{Results}
\label{sec:results}

\subsection{Relativistic corrections}
 
The left panel of Fig.~2 shows the combined single-telescope and interferometric astrometry of the 1992-2018 sky-projected orbital motion of S2, where the zero point is the position of the central mass and of Sgr\,A*. All NACO points were corrected for a zero-point offset and drift in R.A./Dec., which are obtained from the orbit fit. The bottom right panel zooms into the 2018 section of the orbit around pericentre measured with GRAVITY. The zoom demonstrates the hundred-fold improvement of astrometry between SHARP in the 1990s (${\approx}\,4\,\mathrm{mas}$ precision) and NACO in the 2000s (${\approx}\,0.5\,\mathrm{mas}$) to GRAVITY in 2018 (as small as ${\approx}\,30\,\mu\mathrm{as}$). While the motion on the sky of S2 could be detected with NACO over a month, the GRAVITY observations detect the motion of the star from day to day. The upper right panel of Fig.~\ref{fig:fig2} displays the radial velocity measurements with SINFONI at the VLT and NIRC2 at Keck in the 1992-2018 period.

At pericentre $R_{\rm peri}$, S2 moves with a total space velocity of ${\approx}\,7650\,$km/s, or $\beta\!=\!v/c\!=\!2.55\!\times\!10^{-2}$. This means that the first-order parameterised post-Newtonian correction terms (PPN(1)), due to Special and General Relativity, beyond the orbital Doppler and Rømer effects, are within reach of current measurement precision, PPN(1) $\sim \beta^2 \sim (R_S / R_{\rm peri}) \sim 6.5\!\times\!10^{-4}$. These terms can be parameterised spectroscopically as (e.g. \citealt{1973grav.book.....M, 2005PhR...419...65A, 2006ApJ...639L..21Z}).
\begin{equation}
\label{eq:ppn}
  z = \frac{\Delta \lambda}{\lambda} = B_0 + B_{0.5}\beta + B_1\beta^2 + {\cal O}(\beta^3),
\end{equation}
where the PPN$(1)_z$ term $B_1\!=\!B_{1,tD}\!+\!B_{1,gr}$, with $B_{1,tD}\!=\!B_{1,gr}\!=\!0.5$, and $\beta^2\!=\![R_s(1\!+\!e)]/[2a(1-e)]\!=\!6.51\!\times\!10^{-4}$ for S2. Here $a$ is the semi-major axis and $e$ is the eccentricity of the S2 orbit. $B_{0.5}\beta$ is the Newtonian Doppler shift.
 
Eq.\,(\ref{eq:ppn}) indicates that PPN$(1)_z$ consists in equal terms of the special relativistic transverse Doppler effect ($B_{1,tD}$) and the general relativistic gravitational redshift ($B_{1,gr}$), totalling ${\approx}\,200\,$km/s redshift at pericentre, while at apocentre, it amounts to only {6\,}km/s. If the total orbital redshift $z_{\rm tot}$ is separated into a Newtonian/Kepler part $z_{\rm K}$ and a GR correction, one can write \mbox{$z_{\rm tot}\!=\!z_{\rm K}\!+\!f\ (z_{\rm GR}\!-\!z_{\rm K})$}, where $f$ is zero for purely Newtonian physics and unity for GR. In the following we show the residuals $\Delta z\!=\!z_{\rm GR}\!-\!z_{\rm K}.$ The Keplerian part of the orbit is at $\Delta z\!=\!0$, and the PPN$(1)_z$ corrections appear as an excess.

\subsection{Analysis with prior Kepler orbit}

\begin{figure}[t!]
\vspace{-2.5cm}
\hspace{-1.3cm}
\includegraphics[width=18cm]{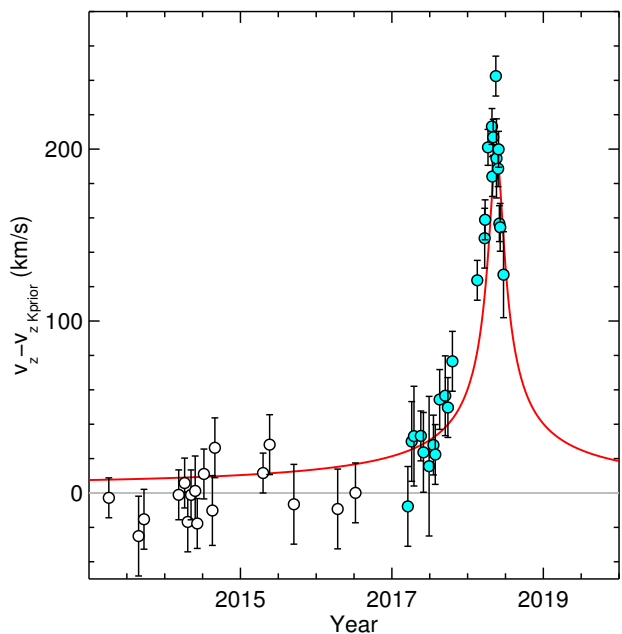}
\vspace{-3.5cm}
\caption{Residual velocity $c \Delta z\!=\!c (z_{\rm GR}\!-\!z_{\rm K})$  for the best fitting prior Keplerian K$_{\rm prior}$ ($f\!=\!0$, grey) and the same orbit with $f\!=\!1$ (red GR$_{\rm prior}$). K$_{\rm prior}$ was constructed from all 1992-2018 astrometric data with NACO \& GRAVITY and the SINFONI data between 2004 and 2016 (open black circles). The 2017/2018 SINFONI data points (black circles with cyan shading) can then be added to test if the spectroscopic data around pericentre follow K$_{\rm prior}$ or the GR$_{\rm prior}$ predicted from K$_{\rm prior}$. The new data points near and up to pericentre, where the $\beta^2$ effects in radial velocity are expected to be important, fall close to the predicted GR$_{\rm prior}$ curve, and exclude the Keplerian prior orbit.}
\label{fig:fig3}
\end{figure}

We define a prior orbit K$_{\rm prior}$ by excluding those data for which the PPN$(1)_z$ corrections matter. For K$_{\rm prior}$ we use the entire 1992-2018 SHARP/NACO and GRAVITY data and the SINFONI data from 2004 up to the end of 2016. We then obtained K$_{\rm prior}$ as described in \citet{2017ApJ...837...30G}, which requires a simultaneous fit of 13 parameters. The R{\o}mer delay is included in the calculation. The resulting orbit is a modest update of \citet{2017ApJ...837...30G}. Using this as the prior orbit, we then added the radial velocities from 2017 and 2018 (Fig.~\ref{fig:fig3}). The 26 residual 2017/2018 spectroscopic data relative to K$_{\rm prior}$ clearly do not follow the best-fitting Keplerian orbit derived from all previous 51 spectroscopic and 196 positions in the past 26 years (grey line in Fig.~\ref{fig:fig3}), but instead follow the $f\!=\!1$ (i.e. GR(K$_{\rm prior}$)) version of K$_{\rm prior}$ (red line in Fig.~\ref{fig:fig3}). This test is fair: GR-corrections should only be detectable with our measurement errors within $\pm 1\,$year of pericentre.

This {\bf a priori} test demonstrates that the spectroscopic data around the pericenter passage are inconsistent with Newtonian dynamics and consistent with GR. However, both K$_{\rm prior}$ ($\chi_r^2\!=\!21$) and GR(K$_{\rm prior})$ ($\chi_r^2\!=\!8$ ) are poor fits to the data.

\subsection{Posterior Analysis}

\begin{figure*}[t!]
\vspace{-0.6cm}
\centering
\includegraphics[width=17cm]{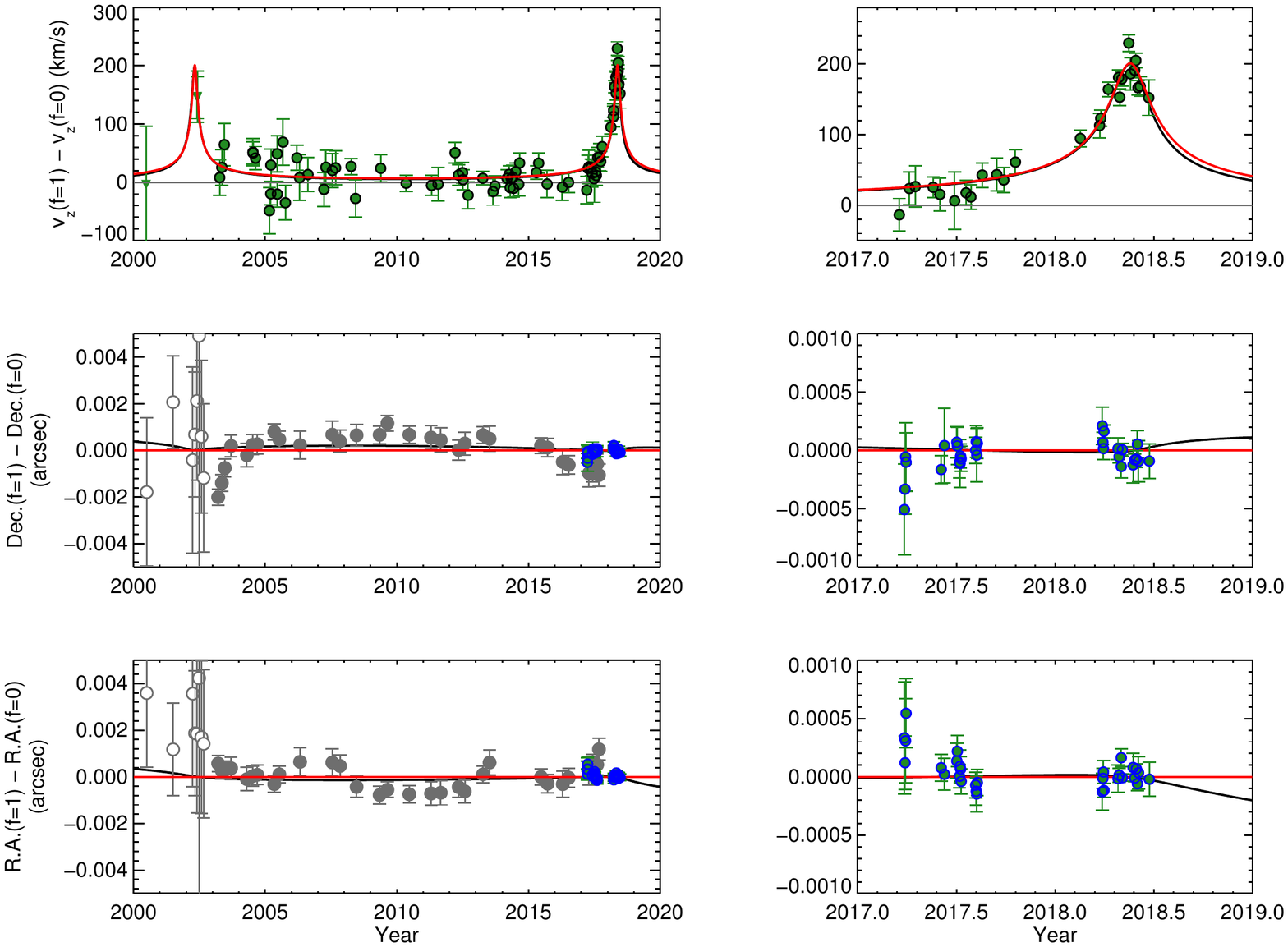} \
\vspace{0cm}
\end{figure*}
\begin{figure*}[h!]
\vspace{-2.5cm}
\centering
\includegraphics[width=18cm]{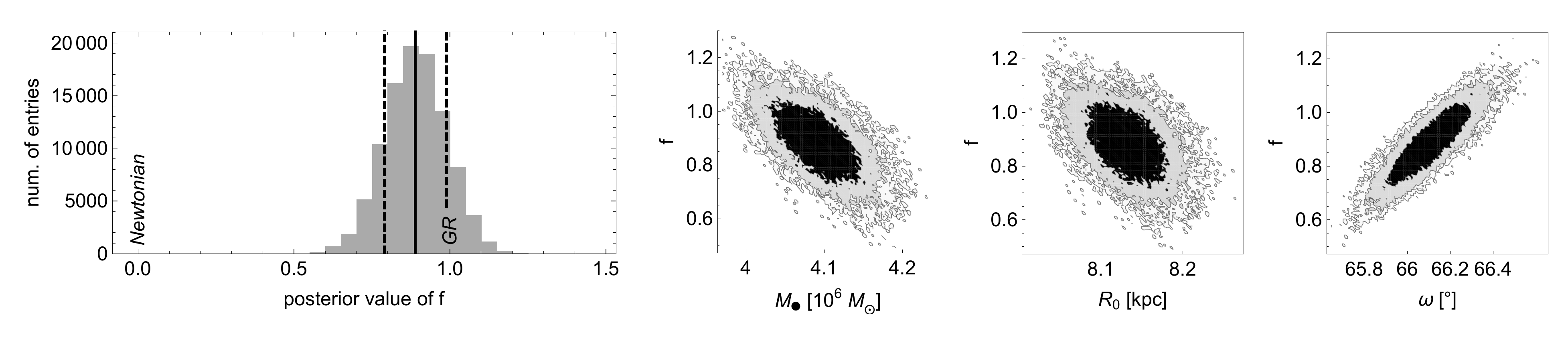}
\vspace{-0.3cm}
\caption{Posterior analysis of all data by fitting for $f$ simultaneously with all other parameters. We plot the residuals in spectroscopy (top, NIRC2, NACO, and SINFONI), Dec. and R.A. (middle two panels, filled grey: NACO; open grey: SHARP; green filled blue: GRAVITY) between the best $f\!=\!1$ fit and the $f\!=\!0$ (Newtonian) part of that fit for the model (red line) and all data. The black curve includes the Schwarzschild precession. Here, we down-sampled the NACO data into 100 equal bins along the orbit to obtain a constant weighting in spatial coverage. With a weight of 0.5 for the NACO data (in order to account for the systematic errors), this yields a $10\,\sigma$ result in favour of GR ($f\!=\!0.90 \pm 0.09$), and  $\chi_r^2\!=\!0.86$. The bottom panel shows the posterior probability distributions for $f$ and its correlation with the mass $M_\bullet$ and distance $R_0$ of the massive black hole, and the argument of periapsis $\omega$. The distributions are compact and all parameters are well determined.}
\label{fig:fig4}
\end{figure*}

Because of the uncertainties in the parameters of K$_{\rm prior}$, in particular, in the strongly correlated mass and distance, a more conservative approach is to determine the best-fit value of the parameter $f$  {\bf a posteriori}, including all data and fitting for the optimum values of all parameters. In carrying out the fitting, it is essential to realise that the inferred measurement uncertainties are dominated by systematic effects, especially when evidence from three or more very different measurement techniques is combined (see Appendix \ref{app:systematics} for a more detailed discussion). In particular the NACO measurements are subject to correlated systematic errors, for example from unrecognised confusion events \citep{2018MNRAS.476.4372P}, which typically last for one year and are comparable in size to the statistical errors. We therefore down-sampled the NACO data into 100 bins with equal path lengths along the projected orbit (Fig. 4, middle) and gave these data in addition a lower weight of 0.5. Depending on exactly which weighting or averaging scheme was chosen, the posterior analysis including all data between 1992 and 2018 yielded $f$ values between $0.85$ and $1.09$. With a weighting of 0.5 of the NACO data, we find $f\!=\!0.90\pm0.09$ (Fig.~\ref{fig:fig4}). GR ($f\!=\!1$) is favoured over pure Newtonian physics ($f\!=\!0$) at the $\approx\!10\,\sigma$ level.

The error on $f$ is derived from the posterior probability distributions (Fig.~\ref{fig:fig4}, bottom) of a Markov chain Monte Carlo (MCMC) analysis. Fig.~\ref{fig:a1} shows the full set of correlation plots and probability distributions for the fit parameters. The distributions are compact and all parameters are well determined. The best-fit values and uncertainties are given Table~\ref{tab:a1}. 

The superb GRAVITY astrometry demonstrably improves the quality of the fits and is crucial for overcoming the source confusion between Sgr\,A* and S2 near pericentre. A minimal detection of PPN$(1)_z$ (Eq.\ (\ref{eq:ppn})) is provided by a combination using only NACO and SINFONI data ($f_{\rm NACO\!+\!SINFONI}\!=\!0.71\pm0.19$, $3.6\,\sigma$), but the inclusion of the GRAVITY data very significantly improves the precision and significance of the fitted parameters: the improvement reaches a factor of 2 to 3. 
 
A still more demanding test is to search for any Keplerian fit to all data and determine whether its goodness of fit is significantly poorer than the goodness of fit of the best-fitting GR-orbit. For linear models the formula presented in \citet{2010arXiv1012.3754A} can be used to estimate the significance. However, the value for the degrees of freedom ($d.o.f.$) is not well defined for non-linear models \citep{2010arXiv1012.3754A}. In our case, we have two models that only differ significantly over a very critical short time-span given the uncertainties in the underlying data. We therefore used the number of those data points as $d.o.f.$ for which the two models predict significant differences. The difference in $\chi^2$ yields a formal significance of $5\,\sigma$ or greater in favour of the relativistic model. 

For further comments on a Bayesian analysis of our data, see Appendix~\ref{app:mcmc}.
  
\section{Discussion}
\label{sec:discussion}

We have reported the first direct detection of the PPN(1) gravitational redshift parameter around the MBH in the Galactic centre from a data set that extends up to and includes the pericentre approach in May 2018. Three different analysis methods of our data suggest that this detection favours the post-Newtonian model with robust significance. Further improvement of our results is expected as our monitoring continues post pericentre. Still, there are reasons to be cautious about the significance of these early results, mainly because of the systematic effects and the validity of our basic assumptions and model. The most important concern probably is that our basic input model (a binary consisting of an MBH and a star with much lower mass) is incomplete. Additional ‘luminous’ and massive objects around S2 and between S2 and Sgr\,A* are unlikely given the spectroscopic and imaging data. Based on the radial velocities of S2, \citet{2018ApJ...854...12C} excluded any companion with $M sin(i)\,<\,1.6\,M_\odot$ for periods up to 150 days, the longest period for which the binary is not subject to tidal break-up.  The GRAVITY imaging data (Fig.~\ref{fig:fig1}) so far do not show any object near Sgr\,A* and S2 brighter than $K\,\approx\,18.5\,$mag, corresponding to a $2 M_\odot$ main-sequence star. However, massive, non-luminous objects, such as stellar black holes, might be present and could affect the orbital dynamics of S2 \citep{2009ApJ...705..361G, 2010PhRvD..81f2002M, 2010MNRAS.409.1146G}. We repeated the exercise by \citet{2017ApJ...837...30G} of testing how much of an extended mass distribution (in form of a Plummer distribution) could still be commensurable with our full new data set. We find that such an extended mass is lower than 0.35 to $1\,$\% of the central mass, depending on the assumed Plummer radius. 
 
The next relativistic correction term we hope to detect is the Schwarzschild precession, which per orbital revolution is 
\begin{equation}
\label{eq:sp}
\Delta \Phi_{\rm per\,orbit} = \frac{3\pi R_S}{a(1 - e^2)}\,{\rm radians} \approx 12'\ {\rm for\ S2}. 
\end{equation}
Since the precession is strongly dependent on distance from the black hole and S2 is on a highly elliptical orbit, the term manifests itself as a kink between the incoming near-Keplerian and the outgoing near-Keplerian orbit. In addition, it leads to a westward drift of all data points around apocentre. The posterior fit of the current data including the Schwarzschild precession yields an $f$-value still closer to GR than without the precession term ($f\!=\!0.94 \pm 0.09$). The chances for robustly detecting the Schwarzschild precession with further observations are very high. GRAVITY will continue to be critical for this second phase of the experiment. Our forecast suggests that we will obtain a $5\,\sigma$ posteriori result with GRAVITY by 2020 \citep{2017A&A...608A..60G}. 

\begin{acknowledgements}

We are very grateful to our funding agencies (MPG, ERC, CNRS, DFG, BMBF, Paris Observatory, Observatoire des Sciences de l’Univers de
Grenoble, and the Funda\c c\~ao para a Ci\^encia e Tecnologia), to ESO and the ESO/Paranal staff, and to the many scientific and technical staff members in our institutions who helped to make NACO, SINFONI, and GRAVITY a reality. S.G., P.P., C.D., N.B., and Y.K. acknowledge support from ERC starting grant No. 306311. F.E. and O.P. acknowledge support from ERC synergy grant No. 610058. We also would like to acknowledge the important theoretical contributions of the late Tal Alexander (Weizmann Institute, Rehovot), whose 2006 paper with Shay Zucker and members of the MPE group encouraged us to pursue this project. Unfortunately, Tal missed seeing the fruits of this effort by only a few weeks.

\end{acknowledgements}

\bibliography{references}

\begin{appendix}

\section{Supplementary material}
  
\subsection{NACO and SINFONI data analysis}
\label{sec:A1}

The data reduction and analysis tools are almost identical to what we used in \citet{2017ApJ...837...30G}, such that in the following section we concentrate on the specific aspects relevant for the excess redshift. We do not use the 'combined' data set in the sense of \citet{2009ApJ...707L.114G}, that is,\ we do not include the astrometric data set presented in \citet{2016ApJ...830...17B}, but use NTT and VLT astrometry only. This facilitates fitting because it removes four fit parameters.

In 2018, the NACO point spread functions (PSF) for S2 and Sgr\,A* overlap, such that confusion affects the S2 positions at the mas level, similar to the data in 2002 \citep{2008ApJ...689.1044G, 2009ApJ...692.1075G}. Fortunately, the GRAVITY astrometry is not affected by this confusion problem and improves the 2017/2018 astrometry to a precision of $30\!-\!150\,\mu$as. In the case of NACO, fainter and so-far unknown stars might be present in the field and could result in $0.5\!-\!1\,$mas positional offsets from undetected source confusion throughout the S2 orbit, which typically lasts for about one year \citep{2018MNRAS.476.4372P}.

Another critical aspect is the precision of radial velocity measurements of S2. It is a $K_s\,{\approx}\,14\,$mag star, for which the HI $n\!=\!7\!-\!4$ recombination line (Brackett-$\gamma$, $\lambda\!=\!2.1661\,\mu$m) and the He-I ($\lambda\!=\!2.112\,\mu$m) line can be detected in absorption at ${>}\,5\,\sigma$ per spectral resolution element in one hour. The wavelength calibration is fine-tuned in each individual exposure by comparing the positions of the atmospheric OH-emission-lines with their expected positions. In the wavelength regime in which the S2 Br$\gamma$ line is currently observed, we use approximately a dozen lines, and the scatter of the OH-lines after the fine-tuning around the expected positions is below 5 km/s, which is smaller than the scatter in the Br$\gamma$ data. We hence estimate that our systematic uncertainty due to the wavelength calibration is 5 km/s. The dominant error term, however, is the correction from residual sky features in the data. Given that S2 changes its radial velocity quickly and that these residuals vary from one observation to the next, we can assume that they essentially act as a random error. In practice, typical $1\,\sigma$ uncertainties of the line centres are $\pm12$ to $\pm20\,$km/s. At some observation epochs, confusion with other stars or extended nebular emission (mainly at low velocities) or atmospheric residuals leads to increased uncertainties. The line shape of S2 might be affected by a stellar wind, although previous analyses suggest that S2 is a main-sequence dwarf with low rotational velocity, which is not expected to have significant mass loss \citep{2008ApJ...672L.119M, 2017ApJ...847..120H}. The stacked spectrum of S2 with a very high signal-to-noise ratio of ${\rm } {\approx}\,200,$ newly obtained during the pericentre passage (March-June 2018), does not show a P-Cygni profile either, which would be indicative of a wind. If a wind component were to introduce a constant bias, it would affect the accuracy, but not the precision by which we measure the redshift of S2. In the fit, this would in turn be absorbed into the motion of the reference system in the direction of the line of sight. If the shape of the spectrum is variable due to the wind, we would obtain a lower precision on the radial velocities. So far, no hints of a variable spectrum of S2 have been seen, and the classification of S2 as a B2.5 main-sequence star argues against spectral variability \citep{2017ApJ...847..120H}. Moreover, we use a cross-correlation with the observed S2 spectrum \citep{2008ApJ...672L.119M, 2017ApJ...847..120H} in addition to a line fit, which would most likely be affected in a different way than the single line. The two ways of determining the radial velocity agree very well, which demonstrates that the line shape of the Br$\gamma$ line does not affect our measurement.

\subsection{GRAVITY observations} 
\label{app:gravity-observations}

The GRAVITY observations were taken at the Very Large Telescope Interferometer in Chile. The instrument coherently combines the light of the four 8m UTs. We chose the most sensitive low spectral resolution mode of GRAVITY \citep{2017A&A...602A..94G}. In this mode, the science spectrum is dispersed across 14 pixels with a spectral resolving power of $R\,{\approx}\,20$. Nearly all data were taken in split polarisation mode, with a Wollaston prism inserted in the optical train and the two linear polarisations recorded independently. 

All four UTs locked their Coud\'e infrared AO (CIAO, \citealt{2016SPIE.9909E..2LS}) module on the brightest source in the field, the red supergiant IRS7  ($m_K\,{\approx}\,6.5\,$mag, distance from Sgr\,A* ${\approx}\,5.5''$). Active field and pupil guiding
was enabled \citep{2018MNRAS.476..459A}. The interferometric observations started with IRS16NW or IRS16NE feeding the fringe-tracker and IRS16C feeding the science channel. These two bright stars ($m_K\,{\approx}\,$10.0\,-\,10.5\,mag, separation from Sgr\,A* ${\approx}\,1''$) were used to find fringes and to zero the optical delay of the science channel. The actual observations of S2 and Sgr\,A* were then made with IRS16C or IRS16NW as fringe-tracking star. Each science exposure consists of 30 frames with an individual integration time of 10s each. The typical observing sequence had five such 5-minute exposures on Sgr\,A*, one exposure on S2, one sky exposure, and one exposure on R2, a moderately bright ($m_K\,{\approx}\,12.1\,$mag, separation ${\approx}\,1.5''$) nearby unresolved giant star, which served as a visibility calibrator. This sequence was repeated several times per night. 

\subsection{GRAVITY data analysis}
\label{app:data-analysis}
  
We used the standard GRAVITY pipeline to process the data \citep{2014SPIE.9146E..2DL, 2017A&A...602A..94G}. Each individual exposure was first sky subtracted, flat fielded, and wavelength calibrated. The data were then reduced based on a pixel-to-visibility matrix (P2VM, \citealt{2007A&A...464...29T}), which represents the instrument transfer function including throughput, coherence, phase shift, and cross-talk information of each individual pixel. In a second step, the science complex visibilities are phase-referenced to the fringe-tracker complex visibilities using the laser metrology and a fiber dispersion model. The observatory transfer function (i.e. coherence loss due to vibrations, uncorrected atmosphere, birefringence, etc.) was calibrated on the nearby unresolved calibrator star R2.

\subsection{Model fitting}
\label{app:model-fitting}

The reported astrometric positions are based on a two-component binary fit to the (squared) visibilities and closure phases. We took into account the flux ratio between S2 and Sgr\,A*, the colour of Sgr\,A*, bandwidth smearing (e.g. \citealt{2013MNRAS.435.2501L}), and a telescope-dependent injection ratio. We developed several independent fitting codes, employing least-squares minimization, MCMC optimization, and a combination of these two techniques. A full mathematical derivation of the models is beyond the scope of this paper. Overall, the results agree very well independent of the optimization technique and the detailed implementation.
 
\subsection{Imaging}
\label{app:imaging}
 
Complementary to the model fitting, we reconstructed images using radio- and optical interferometry imaging tools. By employing different codes, we checked for consistency and robustness.

The radio-interferometry-like imaging was done with the Astronomical Imaging Process System (AIPS, \citealt{2003ASSL..285..109G}) developed at NRAO. For each exposure, we reconstructed the dirty image by a discrete Fourier transform of the complex visibility data. We then extracted preliminary images of S2 using the CLEAN algorithm (e.g. \citealt{1974A&AS...15..417H}) with clean boxes only on the brightest features of the dirty image. After this, we performed phase-based self-calibration of the visibility data with the preliminary S2 model to correct for telescope-based errors. Then we re-ran CLEAN on the self-calibrated data to clean both on S2 and Sgr\,A* iteratively, resulting in one image per exposure. In a last step, we combined individual exposures to obtain the final image for each night. 

We also made use of MiRA2 \citep{2008SPIE.7013E..1IT} and Squeeze \citep{2010SPIE.7734E..2IB}, two optical interferometry imaging codes. These codes fit an image to the data using a least-squares minimization (MiRA2) or MCMC (Squeeze) with some penalty function to account for the sparsity of the data. The advantage of these codes is that they can account for coherence losses due to bandwidth smearing, and that they can also work directly with the robust closure-phases, thereby avoiding the self-calibration described above. Their weakness, however, is that the result depends on the chosen penalty function (priors), and fit convergence can be an issue.
 
\subsection{Treatment of systematic uncertainties}
\label{app:systematics}

As discussed above, all of the major observational input data in this paper (NACO, SINFONI, GRAVITY) are affected by strong and different systematic effects.
NACO positional measurements are affected by obvious or unrecognized confusion events, especially close to Sgr\,A* (typically lasting one year, \citealt{2018MNRAS.476.4372P}) and have to be tied into a long-term reference frame coupled to the radio interferometric frame. Stars common to both frames but spread over tens of arcseconds are used for this alignment, but they in turn require a careful analysis of image distortions and mosaicking shifts \citep{2015MNRAS.453.3234P}. As a result, the S2 positions can at times have systematic uncertainties exceeding $1\,$mas, although the statistical errors are ${\approx}\,0.4\!-\!0.6\,$mas. Depending on the redshift, SINFONI spectroscopic data can be affected more or less strongly by atmospheric sky lines and extended nebular emission in the central cluster. This means that velocity uncertainties can exceed the typical performance in good conditions of $10\!-\!15\,$km/s. GRAVITY astrometry offers by far the best expected positional information ($30\!-\!150\,\mu$as). However, the current accuracy is still limited by systematics and calibration errors, especially if Sgr\,A* is particularly faint, is varying during the integration, or if the S2~-~Sgr\,A* separation is very large and close to the limit of the interferometric and photometric field of view, as was the case in 2017. 

Our approach to account for these effects was to have several team members use different analysis tools for positional fitting and extraction, and to compare and average these results, with the analysis scatter providing an estimate of the systematic effects. Bootstrapping and removal of questionable data sets was performed in all cases. Every new observing epoch will add to the understanding of the systematic effects, and we expect to further improve especially the astrometric accuracy of GRAVITY in the coming years.

\subsection{R{\o}mer effect}
\label{app:roemer}

Both gravitational redshift and transverse Doppler effect are of the order of $\beta^2$. As \citet{2006ApJ...639L..21Z} have pointed out, the classical R{\o}mer delay also needs to be taken into account at that level. Including the light travel time requires solving the equation 
\begin{equation}
\label{eq:a1}
t_{\rm obs} = t_{\rm em} + \frac{x(t_{\rm em})}{c},
\end{equation}
where $x$ denotes the line-of-sight distance, $t_{\rm em}$ the time of emission, and $t_{\rm obs}$ the time of observation. This equation can only be solved iteratively, which is done in some of our codes. However, one can also approximate the correction term by 
\begin{equation}
\label{eq:a2}
\frac{x(t_{\rm em})}{c} {\approx}  \frac{x(t_{\rm obs})}{c}
   \Bigl(1 - \frac{v_x(t_{\rm obs})}{c} \Bigr),  
\end{equation}
which can be evaluated without iteration, and is thus more suitable for a fitting algorithm. For the S2 orbit, the correction term varies by around eight days over the orbit, and the approximation never differs by more than 10 seconds from the exact solution. Some of our fitting codes therefore use the approximation to calculate the R{\o}mer delay. Light bending (lensing and Shapiro time delay) effects on positions (${\approx}\,20\,\mu$as) and velocities (${\approx}\,5\,$km/s) can be neglected at the current precision.

\subsection{Degeneracy between special relativistic effects and gravitational redhift} 
\label{app:degeneracy}

In Eq.\ (\ref{eq:ppn}), the special relativistic transverse Doppler effect and the gravitational redshift are completely degenerate. The degeneracy is broken only by the relative motion between observer and massive black hole, which is mostly due to the solar orbit in the Milky Way. The apparent motion of Sgr\,A* of about $240\,$km/s along the Galactic plane towards the south-west \citep{2004ApJ...616..872R} leads to a correction term that is of the order of $+5\,$km/s at pericentre and $-0.2\,$km/s at apocentre. Earth’s motion around the Sun will contribute a term that is smaller by an order of magnitude. Overall, the effect of the relative motion between observer and Sgr\,A* is too small to break the degeneracy. We therefore use the standard local standard of rest (LSR) correction and accept the complete degeneracy.

\subsection{MCMC analysis}
\label{app:mcmc}

\begin{figure*}[t!]
\centering
\includegraphics[width=\hsize]{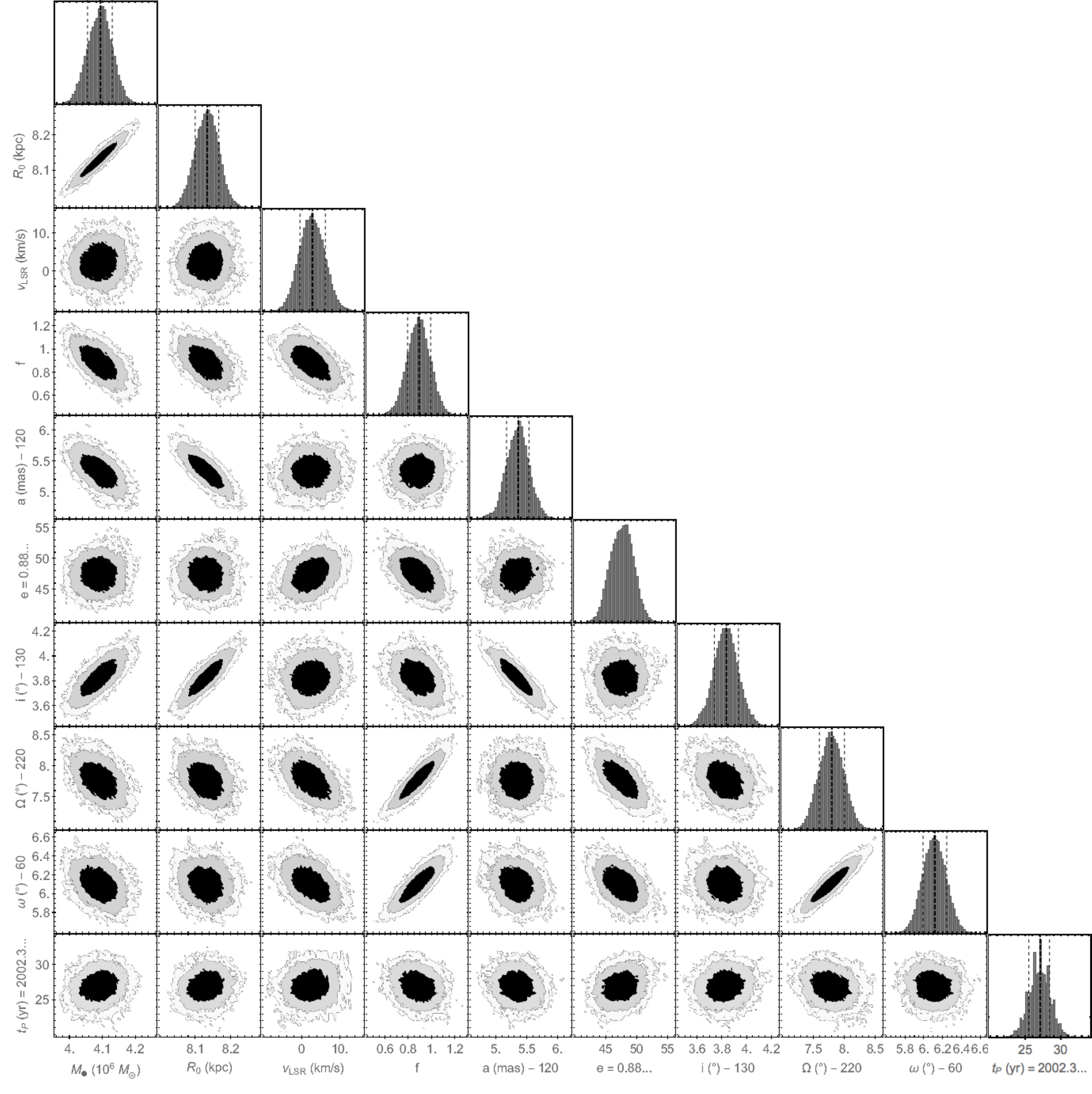}
\caption{Posterior probability distributions obtained from a Markov chain for the 14-parameter fit including $f$ as a free parameter. All parameters are well constrained, in particular also $f$. We have omitted the panels corresponding to the four coordinate system parameters in this figure for clarity.}
\label{fig:a1}
\end{figure*}

As discussed in the main text, we carried out posteriori analyses of all data by fitting 14 parameters characterising the NACO reference frame relative to Sgr\,A* $(x_0, y_0, \dot{x}_0, \dot{y}_0, \dot{z}_0)$, the central mass $M_\bullet$, its distance from the Sun $R_0$, the Keplerian orbit parameters $a,e,i,\Omega,\omega,T_{\rm peri}$, and the parameter $f$ introduced in section \ref{sec:results} to distinguish between Newtonian/Keplerian dynamics ($f\!=\!0$) and the combined special and general relativistic effects up to PPN(1)$_z$ $(f\!=\!1)$. Table~\ref{tab:a1} lists the best-fit solution for the entire data set as treated in Fig.~\ref{fig:fig4} (left without, and right with Schwarzschild precession) for a weight of 0.5 of the down-sampled NACO data. Fig.~\ref{fig:a1} gives the posterior parameter distributions of this fit.

\begin{table}
      \caption[]{Best-fit orbit parameters with and without Schwarzschild precession.}
      \label{tab:a1}
      {\small
      \begin{tabular}{llll}
            \hline
            \noalign{\smallskip}
            Para-& Without  Schwarz-& With Schwarz-& Unit \\
            meter & schild precession & schild precession & \\
            \noalign{\smallskip}
            \hline
            \noalign{\smallskip}
           $f$ & $0.901 \pm 0.090$ & $0.945 \pm 0.090 $ & \\
           $M_\bullet$ & $4.106 \pm 0.034$ & $4.100 \pm 0.034$ & $10^6\,M_\odot$ \\
           $R_0$ & $8127 \pm 31$  & $8122 \pm 31$ & pc \\
           $a$ & $125.38 \pm 0.18$ & $125.40 \pm 0.18$ & mas \\
           $e$ & $0.88473 \pm 0.00018 $ & $0.88466 \pm 0.00018$  & \\
           $i$ & $133.817 \pm 0.093$ & $133.818 \pm 0.093$ & $^\circ$ \\
           $\omega$ & $66.12 \pm 0.12$ & $66.13 \pm 0.12$ & $^\circ$ \\
           $\Omega$ & $227.82 \pm 0.19$ & $227.85 \pm 0.19$ & $^\circ$ \\
           $P$ & $16.0526$ & $16.0518$ & yr \\
           $t_{\rm peri}$ & $2018.37965 \pm 0.00015$ & $2018.37974 \pm 0.00015$ & yr \\
           & $58257.667 \pm 0.054$ & $58257.698 \pm 0.054$ & MJD \\      
           $x_0$ & $-0.88 \pm 0.47$  & $-1.00 \pm 0.47$ & mas \\
           $y_0$ & $-0.97 \pm 0.41$  & $-0.99 \pm 0.41$ & mas \\
           $\dot{x}_0$ & $0.070 \pm 0.031$ & $0.076 \pm 0.031$ & mas/yr \\
           $\dot{y}_0$ & $0.178 \pm 0.030$ & $0.178 \pm 0.030$ & mas/yr \\
           $\dot{z}_0$ & $2.4 \pm 3.0$ & $1.9 \pm 3.0$ & km/s \\
           $\chi_{\rm red}^2$ & 0.86 &  0.86 \\
            \noalign{\smallskip}
            \hline
            \noalign{\smallskip} \noalign{\smallskip}
         \end{tabular} 
         }
         {\footnotesize For the case of Schwarzschild precession, the orbital parameters should be interpreted as the osculating orbital parameters. The argument of periapsis $\omega$ and the time of pericentre passage $t_{\rm peri}$ are given for the epoch of last apocenter in 2010.}
\end{table}

For the purpose of model comparison in a Bayesian framework, we use the fact that the posterior probability distribution is well described by a multivariate Gaussian in both cases ($f\!=\!0$ and $f\!=\!1$). We can thus use the respective peak (best-fit) parameter values and the covariance matrices to approximate the Bayesian evidence integrals \citep{2003itil.book.....M}. The ratio of the so-called Occam factors describing the ratio of the volumes of the two posterior parameter spaces is almost unity, such that the evidence ratio $\rho$ equals the likelihood ratio, i.e. $\rho\,{\approx}\,\Delta \chi^2/2$, which is ${\approx}\,43$ in favour of $f\!=\!1$ (assuming $p(f\!=\!0)\!=\!p(f\!=\!1)$ a priori). The differences in the Bayesian information criteria (BIC) and the Akaike information criteria (AIC) both equal $\Delta\chi^2$ in our case. Given that $\Delta\chi^2\!=\!87,$ we have a ``decisive'' evidence for $f\!=\!1$ when applying Jeffrey's scale. 

We estimated the sampling error on $f$ by means of a bootstrap analysis, during which we randomised the astrometric and spectroscopic data points separately but simultaneously. We refit the bootstrap sample and used the standard deviation of the best-fit values of $f$ as sampling error. We obtain $\Delta f\!=\!0.15$.

\subsection{Lense-Thirring effect}
\label{app:lt}
 
\begin{figure}[t!]
\centering
\includegraphics[width=\hsize]{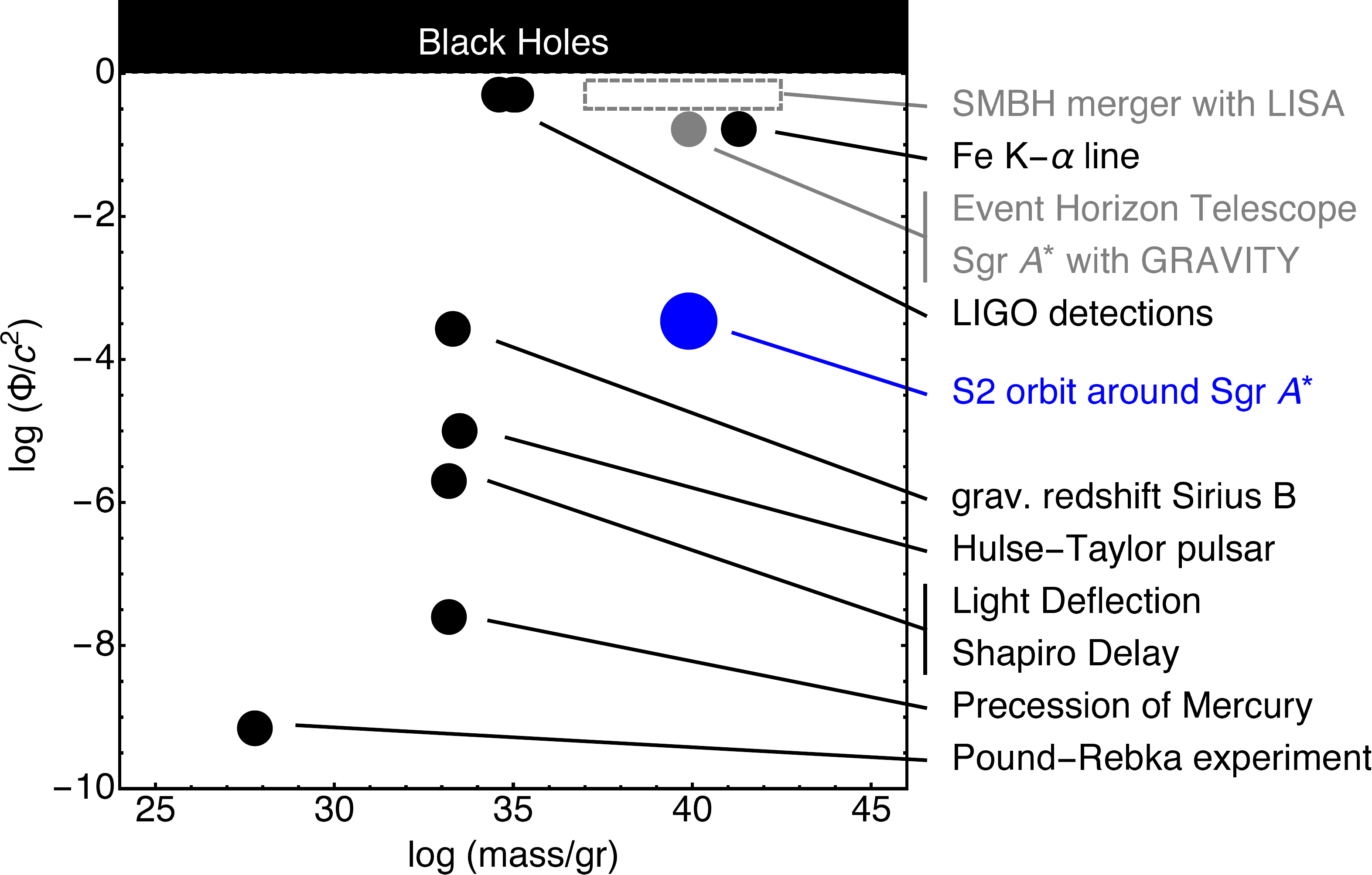}
\caption{Comparison of tests of General Relativity, inspired by \citet{2004AIPC..714...29P}. Shown in black are well-established tests: the \citet{1959PhRvL...3..439P} experiment, the precession of Mercury \citep{1916AnP...354..769E}, light deflection and the Shapiro delay in the solar system, the Hulse-Taylor pulsar \citep{1982ApJ...253..908T}, the gravitational redshift of Sirius B \citep{1971ApJ...169..563G,2005MNRAS.362.1134B}, the LIGO detections \citep{2016ApJ...818L..22A,2016PhRvL.116f1102A, 2016PhRvL.116x1103A}, and the relativistic Fe K$\alpha$ line \citep{1995Natur.375..659T,2000PASP..112.1145F}. Future tests are shown in grey, and this work, which uses the S2 orbit around Sgr\,A*, is shown in blue.}
\label{fig:a2}
\end{figure}

The S2 experiment delivers a valuable confirmation of GR in a so-far unexplored regime at high masses (Fig.~\ref{fig:a2}, adapted from \citealt{2004AIPC..714...29P}). A further goal would be determining the spin of the massive black hole through the combination of frame dragging and quadrupole moment, the so-called Lense-Thirring (LT) precession, of PPN(1.5) order (e.g. \citealt{1973grav.book.....M, 1975ApJ...195L..65B, 2008ApJ...674L..25W}):
\begin{equation}
\label{eq:a3}
\Delta \omega = 2\,\xi \left( \frac{R_S}{a(1 - e^2)}\right)^{3/2}, 
\end{equation}
where $\xi \leq 1$ is the dimensionless spin parameter of a Kerr black hole. For $\xi\!=\!0.5,$ the LT precession of S2 is $9''$, which is clearly not detectable. It is thus necessary to observe stars yet deeper in the potential if the spin of Sgr\,A* is to be measured with orbiting stars. \citet{2018MNRAS.476.3600W} have quantitatively analysed the requirements for detecting the LT-precession on a star inside the S2 orbit. They find that such a star would have to have a combination of semi-major axis $a$ and eccentricity $e$, $a(1\!-\!e^2)^{3/4}/R_S \leq 250$, and a significant detection would require $10\,\mu$as astrometric precision in a campaign over several years.  Based on the K-band luminosity function \citep{2003ApJ...594..812G, 2012A&A...545A..70S, 2018A&A...609A..26G, 2018A&A...609A..27S} and the eccentricity distribution of the S-stars \citep{2017ApJ...837...30G}, \citet{2018MNRAS.476.3600W} estimate a probability of ${\approx}\,10\,$\% for a $K\,{<}\,19\,$mag star to fulfill the above requirement. Still fainter stars would likely be more common. No second star with $K\,{<}\,18.5\,$mag has so far been reliably detected near Sgr\,A* with GRAVITY, which is consistent with the predictions of \citet{2018MNRAS.476.3600W}. We hope for such a detection in the next years, when S2 has moved away from Sgr\,A* and cleared the field of view for fainter objects.
 
\end{appendix}

\end{document}